\documentclass{mn2e}
\usepackage{psfig}
\usepackage{mnras_cite}
\newcommand{\redshift}{\left( \frac{1+z}{7} \right)}   
\newcommand{\redshiftnine}{\left( \frac{1+z}{10} \right)}   

\newcommand{\overdensity}{\left( \frac{\Delta}{3} \right)}

\newcommand{\temperature}{\rm \left( \frac{T}{10^{4} \, K} \right)}       
\newcommand{\metallicity}{\rm \left( \frac{Z}{10^{-2.5} Z_{\odot}}
\right)}
\newcommand{\escape}{\rm \left( \frac{f_{esc}}{0.1} \right)}

\begin{document}

\title[Probing the Dark Ages with Metal Absorption Lines]{Probing the Dark Ages with Metal Absorption Lines} 
\author[Oh] 
{S. Peng Oh\\
Theoretical Astrophysics, Mail Code 130-33, Caltech, Pasadena, CA 91125, USA}

\maketitle

\begin{abstract}
Recent observations of high redshift quasars at $z \sim 6$ have
finally revealed complete Gunn-Peterson absorption. However, this at best
constrains the volume-weighted and mass-weighted neutral fractions to
be $x_{\rm HI}^{\rm V} \ge 10^{-3}$ and $x_{\rm HI}^{\rm M} \ge 10^{-2}$ respectively; stronger constraints are not possible due to the high
optical depth for hydrogen Lyman transitions. Here I suggest certain
metal lines as tracers of the hydrogen neutral fraction. These lines should
cause unsaturated absorption when the IGM is almost fully neutral, if
it is polluted to metallicities ${Z \sim 10^{-3.5}-10^{-2.5}
Z_{\odot}}$. Such a minimal level of metal pollution is inevitable in
the middle to late stages of reionization unless quasars rather than
stars are the dominant source of ionizing photons. The OI line at
$1302$ \AA$\,$is particularly promising: the OI and H ionization
potentials are almost identical, and OI should be in very tight charge
exchange equilibrium with H. The SiII 1260 \AA $\,$transition might also be observable. At
high redshift, overdense regions are the first to be polluted to high metallicity but the last to
remain permanently ionized, due to the short recombination times. Such
regions should produce a fluctuating OI and SiII forest which, if
observed, would indicate large quantities of neutral
hydrogen. 
The OI forest may already be detectable in the 
SDSS $z=6.28$ quasar. If seen in future high-redshift quasars, the OI
and SiII forests will probe the topology of reionization and metal
pollution in the early universe. If in additional the HI optical depth can be measured from
the damping wing of a high-redshift gamma-ray burst, they will yield
a very robust measure of the metallicity of the
high-redshift universe. 
\end{abstract}

\section{Introduction}

Recent spectroscopic observations
\cite{becker01,pent01,dvo01} of high-redshift quasars discovered by
the Sloan Digital Sky Survey (Fan et al 2000, 2001a) have revealed long
gaps in the spectra consistent with zero transmitted flux. This
long-awaited detection of the Gunn-Peterson effect may herald the
observational discovery of the reionization epoch. However, the high oscillator
strength of the hydrogen Ly$\alpha$ transition means that complete
Gunn-Peterson absorption is expected even for a highly ionized
intergalactic medium (IGM). The strongest constraint comes from the
Ly$\beta$ absorption trough, due the weaker (by $\sim 5$) oscillator
strength of Ly$\beta$. From the $z=6.28$ quasar observed
by \scite{becker01}, \scite{fan01b} conclude that at $z \sim 6$, the
lower limits on the mass-weighted and volume-weighted 
neutral hydrogen fraction are $x_{\rm HI}^{\rm M} > 10^{-2}$ and $x_{\rm
HI}^{\rm V} > 10^{-3}$ respectively, larger by almost two orders of
magnitude from $z \sim 4$. Studies interpreting the
observations conclude that the observed
absorption troughs are consistent with the tail end of reionization,
or post-overlap phase after individual HII regions have
merged \cite{barkana,fan01b}. However, due to the
rapid or phase-change like nature of reionization in standard
scenarios \cite{gnedin,raz}, the ``dark ages'' or
pre-overlap phase, when a substantial fraction of the hydrogen in the
universe was neutral, is likely not far off. The spectra of
the $z=6.28$ quasar suggests a very rapid evolution in the effective optical depth and thus the
ionizing radiation field and effective neutral fraction
\cite{fan01b}. This implies that a slightly higher
redshift quasar may indeed lie within the
pre-overlap era.  

Unfortunately, even if such a quasar is discovered, we may not learn anything new about the
pre-reionization epoch. The hydrogen Lyman-series absorption trough
saturates fully for a neutral hydrogen fraction at mean density $x_{\rm HI} \sim 10^{-4}$; because the transmitted flux
declines exponentially with an increasing neutral fraction, we do
not have the power to distinguish between an almost fully neutral IGM and one with only a tiny
neutral fraction. As the IGM becomes almost fully neutral we might observe the red 
damping wing of the Gunn-Peterson trough \cite{jordi}. Unfortunately,
the highly luminous quasars presently observed probably ionize their
surroundings on several Mpc scales; the consequent reduction in
optical depth precludes observation of the red damping wing
\cite{cenhaiman,madaurees}. The only hope of detecting a
damping wing would be to discover objects that ionize only a small
region of the surrounding IGM: either a high-redshift gamma-ray burst
(which has a very short duty cycle) or less luminous quasars or
galaxies (which can be detected through gravitational lensing,
\pcite{ellis}). Alternatively, one might hope to detect the Ly$\alpha$
halo surrounding a high-redshift source as Ly$\alpha$ photons scatter
and redshift in the surrounding neutral IGM \cite{loebryb}. This also
suffers from the difficulty that sources tend to ionize their
surroundings; furthermore, the low surface brightness of the halo implies that
detection is likely only possible with NGST.      

What can be done with present-day technology? Clearly, we need absorption-line probes which are still unsaturated when the
IGM is predominantly neutral. This is possible if the absorbers
are much less abundant than hydrogen or have very small
oscillator strengths. They should have ionization potentials 
similar to that of hydrogen in order to trace the HI
fraction as faithfully as possible. In addition, their absorption lines must lie redward of the
hydrogen Ly$\alpha$ wavelength $\lambda= 1216 \, \AA$, in order to avoid
confusion with the lower-redshift Ly$\alpha$ forest. In this paper, I suggest metal absorption lines as a probe
of the neutral IGM. Metals are a natural probe: for a fully neutral
IGM, $\tau_{\rm H \, Ly\alpha} \sim 10^{5}$ at $z \sim 6$, and while
the oscillator strengths of metal UV/optical transitions ($f \sim 10^{-2}-1$)
are roughly comparable to that of hydrogen Ly$\alpha$, the abundance
by number of metals should be lower by $\sim 10^{-6}-10^{-5}$,
implying $\tau_{\rm metals} \sim 10^{-2}-1$. The most uncertain aspect
of this calculation is the degree to which an IGM polluted by metals can still remain neutral. I argue that because
overdense regions are the first to be polluted with metals but the
last to be permanently ionized (due to the short recombination time), a
scenario of a neutral but metal-polluted IGM is plausible. Nonetheless, because of this uncertainty, a null detection
of absorption will only yield a constraint on the joint
metallicity/ionization state of the IGM. A positive detection,
however, may be our best hope of unveiling an almost fully neutral
IGM with observations of high-redshift quasars in the near
future. In all numerical estimates, I assume a $\Lambda$CDM cosmology
with $(\Omega_M,\Omega_\Lambda,\Omega_b,h, \sigma_{8
h^{-1}},n)=(0.35,0.65,0.04,0.65,0.87,0.96)$.

\section{Can metals be seen at high redshift?}

It is useful to begin by ruling out some promising possibilities. The
best absorption line probes would involve promordial elements, which
are not afflicted with uncertainties associated with the (unknown)
high-redshift metal abundance. The most obvious candidate, hydrogen 21 cm
absorption, has too weak an oscillator strength, the optical depth
across a Hubble volume is $\tau = 4.2 \times
10^{-3} \left< x_{\rm HI} \right> \left( \frac{T_{\rm CMB}}{T_{\rm S}} \right)
\redshift^{1/2}$ (where $T_{\rm S}$ is the spin temperature), and the
observational difficulties in detecting a signal are
formidable\footnote{However, 21 cm {\it emission} from the neutral IGM
might be detectable with the Square Kilometer Array \cite{tozzi}}
\cite{shaver}. Similarly, ${\rm H_{2}}$, which lacks a dipole moment, has too low an oscillator strength. HD does
have a dipole moment and higher oscillator strengths (by a factor of $\sim 1000$)
but insufficient to offset its low primordial abundance $x_{\rm HD} \sim
10^{-7}$. Likewise, the optical
depth of lithium is appreciable only at high redshift $z \sim 500$
\cite{loeb2001}. Deuterium has
roughly the right abundance ($\sim 10^{-5}$) and
oscillator strength (same as H), but its Ly$\alpha$ wavelength lies too close
to the H Ly$\alpha$ transition (offset by only $\sim 82 {\rm km \,
s^{-1}}$) to be useful. HeI has a
meta-stable state 2$^{3}$S state , which becomes populated during the
recombination cascade. Resonance-line absorption from this state
occurs at long wavelengths $\sim 4471, 5876 \, \AA$, which are redward of
hydrogen Ly$\alpha$ as required. However,
due to the relatively short lifetime, $\sim 10^{4}$s, of this state, it
is appreciably populated only in highly dense and significantly
ionized gas: n(2$^{3}$S)/n(He$^{+}$)$= 5.8 \times 10^{-6}
T_{4}^{-1.18}/(1+3110 T_{4}^{-0.5} n_{e}^{-1}) \approx 1.9 \times 10^{-9}
T_{4}^{-0.7} n_{e}$ \cite{clegg}, where $n_{e}$ is the electron
number density in ${\rm cm^{-3}}$. The only possibility left is metal lines. 

The optical depth of a line across a uniform IGM which has been
homogeneously polluted by metals is:
\begin{equation}
\tau=0.16 \, x_{i} \left( \frac{X_{a}}{2.7 \times 10^{-6}} \right)
\left( \frac{f}{0.05} \right) \left( \frac{\lambda}{1302 \AA} \right)
\redshift^{3/2}
\label{eqn:GP_uniform}
\end{equation} 
where $x_{i}$ is the fraction of the metal atoms $a$ at the appropriate
ionization state $i$, $X_{a}=Z/Z_{\odot} \times (n_{a}/n_{H})_{\odot}$
is the abundance by number of metal $a$ relative to hydrogen,
$f$ and $\lambda$ are the oscillator strength and rest wavelength of the
appropriate transition. I adopt
$(n_{\rm C,\odot},n_{\rm 0,\odot},n_{\rm Si,\odot},n_{\rm Fe,\odot})=(3.58,8.49,0.33,0.295)
\times 10^{-4} n_{\rm H,\odot}$ for the solar abundance of
carbon, oxygen, silicon and iron respectively \cite{andersgrav}. For some metals,
particularly Si, this may be an underestimate: the supernovae of
supermassive stars which are thought to form out of very low/zero
metallicity gas overproduce $\alpha$ elements such as Si, S and Ca by factors of a few compared to
solar ratios (\scite{heger}; see Fig. 1 of \scite{ohetal}). In Table 
\ref{table:lines}, I list various lines which I have identified as
promising tracers of the IGM ionization state\footnote{Tables for the atomic
constants are available at http://www.pa.uky.edu/~verner/lines.html}: those with ionization
potentials close to that of hydrogen and strong resonance lines
redward of hydrogen Ly$\alpha$. If the IGM is polluted to a
metallicity of $Z \sim 10^{-2.5} Z_{\odot}$ at $z \sim 6$, the optical
depths are fairly high: there would be a flux decrement of $\sim 5-20
\%$ blueward of these lines, which is certainly detectable in high
signal-to-noise spectra. Note that ions with ionization
potential $I_{i}< 13.6$eV may be rare as the universe is optically
thin to radiation at these wavelengths and these atoms are very easily
ionized to the next stage. Submillimeter fine structure lines of
metals have oscillator strengths which are too low for absorption to
be detectable. Although (as we shall see) metal line absorption
likely produces a fluctuating forest rather than a mean flux
decrement, the relative values of $\tau$ provide a good estimate of
the importance of various transitions. 

\begin{table}
\caption{Metal absorption lines which may potentially be observable in
a nearly neutral IGM. $I_{i}$ is the ionization potential of the ion,
$\lambda$ and $f$ are the absorption wavelength and oscillator
strength, and $\tau$ is the optical
depth of the line across a uniform IGM at $z=6.3$, assuming a metallicity of ${\rm Z=10^{-2.5}
Z_{\odot}}$. Only the species with $I_{i} > 13.6$eV (OI, FeII, SiII) are
likely to be abundant, since the universe is optically thin to
radiation below the Lyman limit.}
\label{table:lines}
\begin{tabular}{lcccc}
Ion & $I_{i}$ (eV) & $\lambda (\AA)$ & $f$ &$ \tau(z=6)$ \\
\hline
FeI & 7.87 & 2484 & 0.557 & 0.11 \\
SiI & 8.1 & 2515 & 0.236 & 0.05 \\
CI   & 11.2 & 1657 & 0.148 & 0.23 \\
OI & 13.6 & 1302 & 0.05 & 0.14\\
FeII & 16.1 & 2383 & 0.3 & 0.05\\ 
SiII & 16.34 & 1260 & 1.18 & 0.13 \\
\end{tabular}
\end{table}
 
OI is a particularly promising tracer. Its ionization potential
$I_{i}=13.618$eV is only $\Delta E=0.19$eV higher than that of hydrogen :
therefore a detection of OI almost certainly signals the presence of
neutral hydrogen. In fact, oxygen should be locked in tight charge
exchange equilibrium with hydrogen, through the processes ${\rm O + H^{+}
\rightarrow O^{+} + H^{o}}$, ${\rm O^{+} + H^{o} \rightarrow O +
H^{+}}$ \cite{osterbrock}. The equilibration timescale is $\sim
1/k_{ce} n_{\rm HI} \sim 1.7 \times 10^{5} x_{\rm HI}\Delta \redshift^{3} $years (where $\Delta$ is
the gas overdensity), much shorter than the Hubble time. Therefore the
OI fraction should be very accurately given by
$\frac{n_{O}}{n_{O^{+}}}=\frac{9}{8}\frac{n_{H^{o}}}{n_{p}} {\rm
exp}\left( \Delta E/k_{B}T \right)$, where ${\rm exp}\left( \Delta
E/k_{B}T \right) \rightarrow 1$ for $k_{B}T \gg \Delta E =
0.19$eV. 

The SiII $1260 \, {\rm \AA}$ line is another promising absorption
feature. Note that SiII has a $1304 \, {\rm \AA}$ transition at almost the same wavelength
as the OI transition, but due to the weak oscillator
strength $f=0.0871$ it has an optical depth only $\sim 7 \%$ that of
OI, unless the Si to O ratio is strongly enhanced. Similarly, the $\lambda=1526
\, {\rm \AA}$ transition of SiII ($f=0.132$) has $\tau \approx 0.11
\tau_{\rm OI}$.   

There are two large uncertainties in the above estimates. The first is the mean
metallicity of the IGM at high redshift. The very short timescale on
which massive stars evolve $\sim 10^{6}-10^{7}$years implies that the
IGM could have been polluted very early. The metal abundance of the quasar
environment inferred from spectra of the $z=6.28$ SDSS quasar is
indistinguishable from that of lower redshift quasars; there is little
or no evolution in the observed supersolar metallicities from $z\sim 6$ to
$z\sim 2$, implying that the first stars around quasars must have
formed at $z > 8$ \cite{pent01}. The mean metallicity of the IGM at
$z \approx 3$ at the lowest observable column densities is $Z \sim
10^{-2.5} Z_{\odot}$ \cite{song97,ellison}. From the spectra of 32 quasars in the redshift
range $2.31-5.86$, \scite{song01} finds no evolution in the mean
universal metallicity of the IGM in the redshift range
$z=1.5-5.5$. The {\it minimum} value she finds at $z=5$ is $ Z >
10^{-3.5} Z_{\odot}$; this is a strictly minimal estimate because it
assumes CIV and SiIV are the dominant ionization stages (which may
no longer be true at higher redshifts as the metagalactic radiation
field drops and gas densities increase); furthermore, lines are substantially
undercounted at the highest redshifts because of high noise levels at
the longest wavelengths. Very plausible theoretical scenarios can be
constructed in which an early generation of stars pollute the
intergalactic medium to mean metallicities $ Z \approx 10^{-2.5}
Z_{\odot}$ with volume filling factors $> 20 \%$ by $z \sim 9$,
without significant hydrodynamic perturbation of the IGM
\cite{madau01}. 

In fact, we emphasize that for reionization to take place, a
significant amount of star formation and thus metal ejection must take
place. \scite{madaushull} find that the energy in the ionizing
continuum released by stars is $\sim 0.2 \%$ of the rest-mass energy
of metals produced, or $\sim 1.8$ MeV per metal baryon; this is fairly
independent of the IMF since the same massive stars which
produce ionizing photons produce metals. A similar
relation holds for supermassive stars $M > 100 M_{\odot}$ thought
to form out of metal-free gas, which explode as pair-instability
supernovae: they eject $\sim$half of their mass as metals
\cite{heger}, and release $1.8 \times 10^{48}$ HI ionizing photons
${\rm s^{-1}} \, {\rm M_{\odot}^{-1}}$ over a lifetime $\sim 3 \times 10^{6}$yrs
\cite{bromm}, which yields $\sim 3.3$ MeV per metal baryon. Using the
\scite{madaushull} relation and assuming $\sim 20 \, {\rm eV}$ per HI ionizing
photon, we can write:
\begin{equation}
n_{\gamma} = 0.7 { \left( \frac{\bar{Z}}{10^{-2.5} Z_{\odot}}
\right)} \escape + n_{\gamma}^{\rm QSO}
\label{eqn:num_photons}
\end{equation}  
where $n_{\gamma}$ is the number of ionizing photons per baryon in the
universe, $\bar{Z}$ is the mean metallicity of the universe, $f_{\rm
esc}$ is the escape fraction of ionizing photons from their host
halos, and $n_{\rm QSO}$ is the contribution from quasars (which
produce ionizing photons but no metals). The escape fraction $f_{esc}$
is highly uncertain, with estimates ranging from $\sim 5 \%$ in the
local universe \cite{leitherer,doveshull} to as high as $\sim 50 \%$
in highly luminous Lyman break galaxies \cite{steideletal}. However,
even for $f_{esc} \sim 100 \%$, the mean metallicity should be
reasonably high toward the tail end of the reionization process,
$\bar{Z} \sim 10^{-3.5} Z_{\odot}$. Another uncertainty is the filling
factor of metal-polluted regions $f_{Z}$. We show in \S
\ref{section:OI_forest} that most reasonable values of $f_{Z}$ should
give rise to an absorption signal. Significant retention of metals by
their host halos is unlikely due to the shallow potential wells
predominant at these early epochs; in addition metal-enriched material
is much more easily ejected from halos than the ambient gas
\cite{maclow}. The only case where the universe
can be reionized without significant co-production of metals is if
quasars are the dominant ionizing source. Again this is highly
uncertain, but note that the comoving emissivity of quasars at $z =5$,
as inferred from the quasar luminosity function, is insufficient to
keep the universe reionized at $z=5$ by an order of magnitude
\cite{madau:rad_transfer,fan01a}; the density of quasars at high redshift is
also constrained by the lack of faint red unresolved objects in the
Hubble Deep Field \cite{haiman:qso_density}. Except for the very early
stages of reionization (which are unlikely to be accessible
with present-day instruments, in any case), the IGM is likely to be
polluted to sufficiently high metallicity to make metal-line
absorption studies feasible.  

The second, much larger uncertainty is whether regions which are
pre-enriched with metals can still remain neutral. As previously
noted, the filling factor of ionized regions should be considerable
once the IGM is polluted up to metallicities $Z \sim
10^{-3.5}-10^{-2.5} Z_{\odot}$, unless the escape fraction is very
small $f_{esc} < 1 \%$, and it is possible that all metal-polluted
regions will also be ionized. Indeed, for $f_{esc} > {\rm few} \, \%$,
the typical size of HII regions at $z\sim 9$ will be greater than that
of the metal-laden supernovae-driven superbubble \cite{madau01}; an
ionization front precedes the metal-pollution front. Even if the
ionizing-photon escape fraction is extremely small, the metal pollution front
is likely to collisionally ionize the IGM by shock heating it to $T >
10^{4.5}$K, due to the high speed of the expanding superbubble.   

However, it is important to realize that hydrogen recombination times at high
redshift, $t_{\rm rec} \approx  3 \times 10^{8} x_{e}^{-1} \Delta^{-1}
\redshiftnine^{-3} \temperature^{0.7}$yrs, are short compared to the
Hubble time, $t_{H}=9 \times 10^{8} \redshiftnine^{-1.5}$yrs, so the
ionization fraction, $x_{e} = 1/(1+(t/t_{\rm rec})) \approx
t_{\rm rec}/t_{H}= 0.3 \Delta^{-1} \redshiftnine^{-1.5}
\temperature^{0.7}$, and the gas could become $\sim 70 \%$ neutral. The
lifetime of sources is likely to be short: the lifetime
of massive stars is $t_{MS} \sim 10^{6}-10^{7}$yrs, and the duty cycle
of quasars is probably of order the Eddington timescale $\sim
10^{7}$yrs (such a lifetime is consistent with current observations of
quasars; see \scite{blandford} and references therein). Early reionization
in the pre-overlap era is likely a highly stochastic process in which
regions of the IGM are ionized, polluted with metals, and then
recombine and become largely neutral until another source lights
up. While early reionization is temporary and the ionization state of
any given region of the IGM fluctuates, metal pollution is a permanent
process and the metallicity of the IGM rises monotonically. Note also that
although high-density filaments may initially be collisionally ionized
by the accretion shock during gravitational collapse, the gas will
eventually cool and recombine.   

In particular, overdense regions are the first to be polluted up to high
metallities (due to their proximity to sites of star formation) but
they are the last to remain permanently ionized (due to the short recombination
times). From numerical simulations,
\scite{cenostriker} find that metallicity depends very strongly on
local density: at every epoch, higher-density regions have much higher
metallicities than lower-density regions. In fact, the highest-density regions quickly saturate at near-solar metallicities early
on. These results are in much better agreement with observations than
scenarios in which metal pollution is uniform. \scite{jordietal} point out that in an inhomogeneous universe
reionization should begin in voids and gradually penetrate into overdense
regions; the regions of highest density are the last to be
reionized. This picture is strongly substantiated in numerical
simulations of reionization \cite{gnedin}. These arguments suggest that a line of sight to a high-redshift quasar will
intersect regions at or above the mean density which are largely
neutral but nonetheless polluted with metals. Such overdense regions are the
most likely sites to produce the metal absorption lines we seek.     

The temperature of metal-polluted gas is likely to be $\sim
10^{4}-10^{5}$K  \cite{madau01}. It can be shock heated up to $\sim
10^{7}$K by the expanding superbubble, but cools rapidly by Compton
cooling off the CMB on a timescale $t_{\rm comp}=2.3 \times 10^{8}
\redshiftnine^{-4}$yrs. Below $T \sim 10^{4}$K the gas recombines and
Compton cooling (as well as hydrogen line cooling) is no longer
effective; the gas then cools only by adiabatic expansion on the
Hubble expansion timescale. In highly overdense and metal-polluted
regions metal-lines cooling will become important, but these
correspond to collapsed halos which are in any case unstable to star formation. $T \sim 10^{4}$K is also the equilibrium temperature if a photoionizing background is present.  

\section{The OI forest}
\label{section:OI_forest}

\begin{figure}
\psfig{file=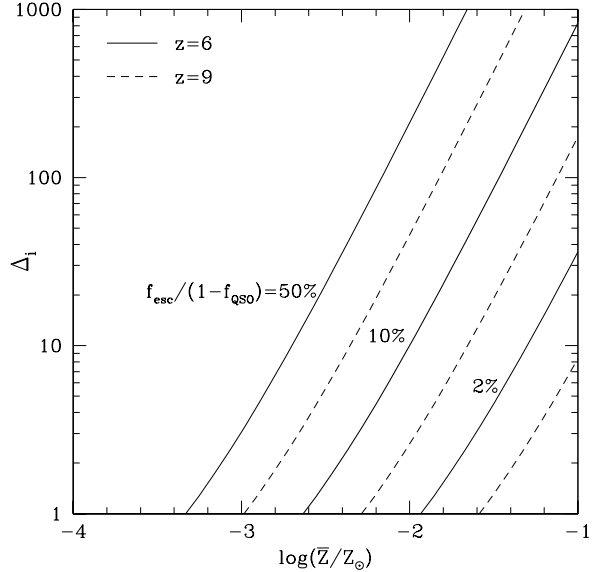,width=80mm}
\caption{Overdensity $\Delta_{i}$ up to which the universe is ionized in the
post-overlap era, as a function of the mean metallicity of the
universe $\bar{Z}$, assuming a (fairly robust) relation between the
metals and ionizing photons produced by massive stars. Solid lines are
for $z=6$, dashed lines for $z=9$. The curves are
plotted for different values of $f_{\rm esc}/(1-f_{\rm QSO})$ where
$f_{\rm esc}$ is the escape fraction of ionizing photons from
star-forming halos, and $f_{\rm QSO}$ is the fractional contribution of
quasars to the ionizing background. The relation between $\Delta_{i}$ and
$\bar{Z}$ shown here is used in Figure \ref{fig:post_overlap}.} 
\label{fig:overdensity_zbar}
\end{figure}

\subsection{A simple model for gas clumping and metal pollution}

Let us now quantify the effects of gas clumping in the
IGM. \scite{jordietal} find the following to be a good fit to the
probability distribution by volume of gas overdensities $\Delta$ seen in the LCDM numerical simulations of \scite{jordi1996}:
\begin{equation}
P_{V}(\Delta) d \Delta = A {\rm exp} \left[ -
\frac{(\Delta^{-2/3}-C_{o})^{2}} {2 (2 \delta_{0}/3)^{2}} \right]
\Delta^{-\beta} d\Delta
\end{equation}
where they tabulate values for $A,\beta,C_{o},\delta_{o}$ at different
redshifts z=2, 3, 4 and 6. One can extrapolate their results to higher
redshifts by using $\delta_{o}=7.61/(1+z)$ (which fits their results
to better than $1 \%$), assuming $\beta=2.5$ (corresponding to an
isothermal slope for high-density halos), and fixing $A,C_{o}$ by
requiring the total mass and volume to be normalized to unity. Their
fit is valid if the gas is smoothed on the Jeans scale for a gas
temperature $T \sim 10^{4}$K; we have argued that high-redshift metal-polluted gas should indeed be at approximately this temperature. The
fraction of baryons above a given overdensity $\Delta_{i}$ by volume
and by mass is then given by $f_{\rm V}(\Delta_{i})=
\int_{\Delta_{i}}^{\infty} P_{\rm V}(\Delta) d\Delta$ and
$f_{\rm M}(\Delta_{i})= \int_{\Delta_{i}}^{\infty} \Delta P_{\rm
V}(\Delta) d\Delta$ respectively. 

In order to compute the optical depth of the IGM to metal line
absorption we need to specify two unknown functions, the metallicity
$Z(\Delta,z)$ and ionization fraction $x_{i}(\Delta,z)$ of the IGM as
a function of overdensity and redshift, but only in the combination $Y(\Delta,z) \equiv x_{i} (\Delta,z)
Z(\Delta,z)$. We can make progress by making some simplifying assumptions. The
growth of metallicity has two free parameters: the mean metallicity of
the universe $\bar{Z}$ (a proxy for the total amount of star
formation), and the volume filling factor of metal-polluted regions
$f_{Z}$. These two parameters are obviously inter-related, but due to
the large uncertainties we treat them as independent free
parameters, with an upper bound on the filling fraction $f_{Z} < 0.2
{\rm \left(\bar{Z}/10^{-2.5} Z_{\odot} \right)^{3/5}}$ (where
the normalization is based on the model of \scite{madau01}, and the
exponent mimics the energy dependence of the adiabatic Taylor-Sedov
solution). Filling factors as low as $\sim 1\%$ are possible if the
metal-enriched ejecta have magnetic fields which resist mixing with
the IGM; metal-polluted regions could be restricted to magnetized
``streaks'' which are then sheared and distorted by subsequent gravitational
clustering \cite{madau01}. The growth of reionization has two free parameters, the filling factor $Q$ of
ionized regions, and the overdensity up to which gas is ionized
$\Delta_{i}$. Again these parameters are obviously related, but we can
make the approximation that they are decoupled, with the following
argument. The early stages of reionization are characterized by
reionization of the voids ($\Delta_{i} < 1$) and growth of the filling
factor of ionized regions $Q$. However, at some point overlap occurs
($Q \approx 1$), and overdense regions start to be reionized, and
$\Delta_{i}$ grows (this hinges on the fact that high-density regions
only occupy a small fraction of the volume). As noted by
\scite{jordietal}, the value of $\Delta_{i}^{\rm overlap}$ at which $Q \approx 1$ depends on the nature of the ionizing
sources: for dim but numerous sources $\Delta_{i}^{\rm overlap} \sim 1$,
whereas for bright but rare sources $\Delta_{i}^{\rm overlap}$ is
larger, since higher density regions have to be ionized before
percolation can occur. We therefore treat it as a free parameter. 

The progress of reionization and the growth in metallicity
$\bar{Z}$ are coupled via equation (\ref{eqn:num_photons}). In
particular, the relation between $(Q,\Delta_{i})$ and $\bar{Z}$
depends on the escape fraction of ionizing photons $f_{esc}$ and the
relative contribution of QSOs to reionization $f_{\rm QSO} \equiv
n_{\gamma}^{\rm QSO}/n_{\gamma}$. In the early stages of reionization,
recombinations are unimportant and $Q \propto n_{\gamma} \propto
\bar{Z}$. In the late stages when recombinations dominate the
consumption of ionizing photons, we can relate $\bar{Z}$ and the
overdensity up to which gas is ionized $\Delta_{i}$ by equating the number of ionizing photons per baryon
with the mean number of recombinations per baryon in a Hubble time of
the ionized gas:
\begin{equation}
n_{\gamma}(\bar{Z})=t_{H} \alpha \bar{n} C_{\rm HII}(\Delta_{i})
\label{eqn:clumping}
\end{equation}
where $C_{\rm HII}(\Delta_{i})=\int_{0}^{\Delta_{i}} \Delta^{2}
P_{V}(\Delta) d\Delta$ is the clumping factor of ionized gas. This
assumes that most star formation and hence metal pollution occured
during the last Hubble time. The
relationship between $\Delta_{i}$ and $\bar{Z}$ is shown in Fig
\ref{fig:overdensity_zbar}, for different values of $f_{\rm
esc}/(1-f_{\rm QSO})$: the smaller the value of this parameter, the
larger the amount of star formation and thus metal pollution $\bar{Z}$
needed to keep the universe ionized.

With this simple picture we can make an ansatz for the evolution of
$Y(\Delta,z)$. I assume that metal pollution begins in the most
overdense regions, while reionization begins in the most underdense
regions. At any given epoch, I assume $Z \approx Z_{\rm crit}$ for $\Delta >
\Delta_{i}^{f_{Z}}$ and $Z \approx 0$ otherwise. Similarly, I assume the
neutral fraction $x_{HI} \approx 1$ for $\Delta > \Delta_{i}^{HI}$ and
$x_{HI} \approx 0$ otherwise. The value of $\Delta_{i}^{f_{Z}}$ is
given by the implicit equation $f_{\rm V}(\Delta_{i})= f_{Z}$, while
$Z_{\rm crit}=\bar{Z}/f_{M}(\Delta_{i}^{f_{Z}})$. The value of
$\Delta_{i}^{HI}$ is $\Delta_{i}^{HI} < 1$ during the pre-overlap
phase and is given by equation (\ref{eqn:clumping}) in terms of
$\bar{Z},f_{\rm esc},f_{\rm QSO}$ in the post-overlap phase, when
recombinations are important. There are two limits to consider. In the
early stages of reionization, when $\Delta_{i}^{HI} <
\Delta_{i}^{f_{Z}}$, the high-density regions where metals reside are
largely neutral, $x_{i} \sim 1$. The growth in $Y$ with time is
dominated by the increase in metallicity. This epoch can therefore be
characterized by the two parameters $(\bar{Z},f_{Z})$.  In the late
(post-overlap) stages of reionization, when $\Delta_{i}^{HI} >
\Delta_{i}^{f_{Z}}$, the evolution in $Y$ is dominated by the
evolution of $x_{i}$, when increasingly dense (and metal-polluted)
regions become ionized. In this regime the model can be specified in terms of the parameters
$(\bar{Z},f_{z},f_{\rm esc}/(1-f_{\rm QSO}))$, and $\Delta_{i}^{HI}$
can be computed from equation (\ref{eqn:clumping}). The transition
between these two regimes occurs when the volume filling fraction of
metal polluted regions and neutral regions are comparable, $f_{Z} \sim 1-Q$. Since we expect
$f_{Z} < 0.2$ (metal pollution should not be effective in
voids, which occupy most of the volume), the transition regime occurs
roughly at the point of overlap, when $Q \rightarrow 1$. The
transition occurs very quickly: $1-Q$ evolves extremely rapidly at the point of
overlap, when the mean free path of ionizing photons rises on a very
short timescale (of order the light travel time across an HII region)
and the ionizing background increases dramatically. By contrast, $Z$
evolves on the timescale for structure formation or $t_{H}$. In fact,
$\dot{Z}$ may drop at the point of overlap since the sudden rise in
the IGM temperature and Jeans mass could cause a drop in the comoving star-formation rate \cite{BarkanaLoeb}.

In summary, for most of the gas, the parameter $Y \equiv x_{i} Z$ rises
in the early stages of reionization as the metallicity grows, peaks at
an epoch roughly correponding to the overlap epoch, and then falls
as the gas is reionized. Note that for higher $\Delta$, the peak value
of $Y$ is higher and occurs at progressively later epochs, since gas at higher
overdensities is only reionized at later times, and so there is a
longer time interval for metal pollution to take place. We now examine
the observational predictions of this simple model.

\begin{figure}
\psfig{file=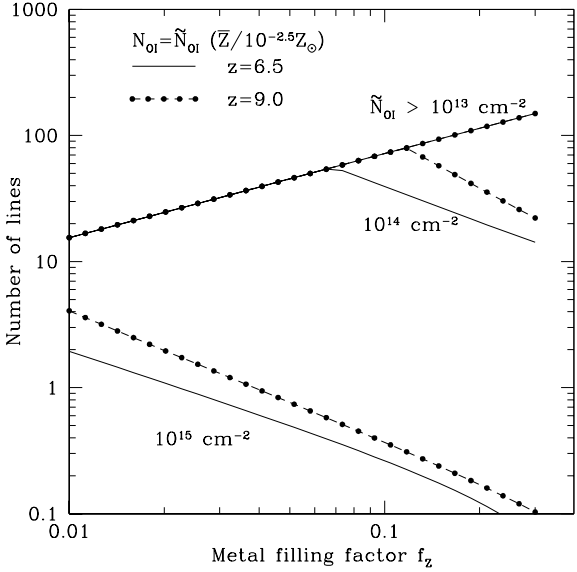,width=80mm}
\caption{Number of OI lines above a given column density $N_{OI}$
observable in the pre-overlap phase, when metal-polluted high density
regions are still largely neutral, as a function of the volume filling
factor of metals $f_{Z}$. Solid lines are for $z=6.5$, dashed lines for
$z=9$. The column density scales directly with the
assumed mean metallicity $\bar{Z}$. The slope of the relation depends
on whether $\Delta_{i}^{N_{OI}} < \Delta_{i}^{f_{Z}}$ (in which case
$N_{\rm lines}$ increases with $f_{Z}$) or $\Delta_{i}^{N_{OI}} >
\Delta_{i}^{f_{Z}}$ ($N_{\rm lines}$ decreases with $f_{Z}$). See text
for details. The number of observable SiII lines with comparable
equivalent widths is roughly half that of OI.}
\label{fig:pre_overlap}
\end{figure}

\subsection{Observational Predictions}
\label{subsection:observations}

We begin by computing the mean Gunn-Peterson absorption in
a clumpy and inhomogeneously polluted universe. The optical depth due
to regions of overdensity $\Delta$ is given by $\tau(\Delta,z)=\Delta
\frac{Z(\Delta,z)}{\bar{Z}} \frac{x_{i}(\Delta,z)}{\bar{x}_{i}}
\tau_{o}(\bar{Z},\bar{x}_{i},z)$, where
$\tau_{o}$ is the optical depth of the line in a uniform IGM which is
uniformly polluted to a metallicity $\bar{Z}$ and has a mean
ionization fraction $\bar{x}_{i}$. The mean metal-line Gunn-Peterson
absorption $\cal{A}$ due to gas with $\Delta > \Delta_{\rm crit}$ is
then given by:
\begin{equation}
{\cal A}(\Delta_{\rm crit}) =\left< 1 -e^{-\tau} \right> =
\int_{\Delta_{\rm crit}}^{\infty} (1-e^{\tau(\Delta)}) P(\Delta)
d\Delta. 
\end{equation} 
Note that in general ${\cal A}(\Delta_{\rm crit})$ is smaller than
${\cal A}$ for a uniform IGM. For instance, for $\Delta_{\rm crit} \sim 3$ at
$z=6$ and $\bar{Z}=10^{-2.5} Z_{\odot}$, we have $\tau_{\rm eff} \approx
0.02 x_{i}$, rather than $\tau \approx 0.14 x_{i}$ for a uniform
IGM. $\tau_{\rm eff}$ falls exponentially with increasing $\Delta_{i}$,
due to the exponentially small fraction of baryons at high overdensities.  

The tight charge-exchange equilibrium between OI and HI implies that there is a
direct relation between their effective optical depths, {\it
independent} of gas clumping or the nature of the ionizing radiation field:
\begin{equation}
\tau^{\rm eff}_{\rm OI}= 1.1 \times 10^{-6} {\left( \frac{\left< Z
\right>}{10^{-2} Z_{\odot}} \right)} \tau^{\rm eff}_{\rm HI},
\end{equation}
where ${\left< Z \right>}$ is the HI column-density weighted
metallicity of the universe (as opposed to the mean metallicity
$\bar{Z}$; in general ${\left< Z \right>} > {\rm \bar{Z}}$). Therefore,
if we could measure both $\tau^{\rm eff}_{\rm OI}$ and $\tau^{\rm eff}_{\rm
HI}$, we can obtain a robust and relatively model-independent measure
of the metallicity of the high-redshift universe. Such a fortuitous
occasion might arise if we could observe a
high-redshift gamma-ray burst, which does not exhibit a strong
proximity effect, and therefore allows measurement of $\tau^{\rm
eff}_{\rm HI}$ by measuring the shape of the damping wing. The transmitted flux recovers
its full value at $\Delta \lambda /\lambda \sim 0.1$ redward of the
damping wing  \cite{jordi}, so a clean separation of the contribution
of OI absorption at $1302(1+z_{s}) \, {\rm \AA}$ ($\Delta\lambda/\lambda
\sim 0.07$) might be possible, particularly if the shape of the
damping wing is well-constrained. Although OI absorption will probably
produce a fluctuating forest, $\tau_{\rm OI}^{\rm eff}$ can be obtained by
smoothing the spectrum. The ability to measure a mean OI decrement of
a few percent depends on the accuracy to which sky lines and absorption
from other sources can be ruled out (see discussion at end of
section). In the absence of a measurement of $\tau_{\rm HI}^{\rm
eff}$, a conservative lower bound on the metallicity of the
high-redshift universe can still be placed from equation
(\ref{eqn:GP_uniform}) by assuming the IGM to be fully neutral, $x_{i}
\sim 1$, and uniform (since $\left< \tau_{\rm OI} \right> > \tau_{\rm OI}^{\rm eff}$).    

Metals in the high-redshift IGM will probably not produce a Gunn-Peterson-like
absorption trough but rather a forest of metal lines
which$--$particularly in the case of OI$--$will provide a snapshot of
neutral regions along the line of sight. \scite{schaye} shows that many properties of the Ly$\alpha$
forest can be understood by associating the characteristic lengthscale
of absorbers with the local Jeans length. This allows us to associate
a column density for an ion $i$ with a given overdensity $\Delta$:
\begin{eqnarray}
N_{i}= 6.2 \times 10^{13} \, {\rm cm^{-2}} \, x_{i} \left( \frac{X_{a}}{2.7 \times 10^{-6}}
  \right) \redshift^{1/2} \\ \nonumber
\times \temperature^{1/2} \overdensity^{1/2} 
\end{eqnarray} 
where $x_{i}$ and $X_{a}$ are the ionization fraction and metal number
abundance. This corresponds to an equivalent width $\left(
\frac{W_{\lambda}}{\lambda} \right)=5.8 \times 10^{-5} \left(
\frac{N_{i}}{10^{14} \, {\rm cm^{-2}}} \right) \left(
\frac{f}{0.05} \right) \left( \frac{\lambda}{1302 \AA} \right)$, where
I have used the fact that lines will always be on the linear portion
of the curve of growth. This gives an observed equivalent width
\begin{equation}
W_{\lambda} \approx 0.53 \, {\rm \AA} \left( \frac{N_{\rm OI}}{10^{14} \, {\rm cm^{-2}}}
\right) \redshift,
\end{equation}
which is certainly detectable with extended
integration on Keck. At a given overdensity $\Delta$ the OI $1302 \, {\rm \AA} \,$ and SiII
$1260{\rm \AA} \,$ equivalent widths are roughly equal; the increased oscillator
strength of the SiII line compensates for its reduced abundance. 
Since on average $W_{i}$ increases monotonically with $\Delta$, for any given $W_{\rm i}$, there exists some $\Delta_{\rm i}$ such that $\Delta > \Delta_{i}
\Rightarrow W > W_{\rm i}$. The mean spacing between lines
with $W > W_{\rm i}$ can be estimated by the mean separation
between contours of overdensity $\Delta_{i}$ in the
universe. This can be estimated as \cite{jordietal}:
\begin{equation}
\lambda_{i}=\lambda_{o} \left[ 1 - F_{V}(\Delta_{i}) \right]^{-2/3},
\end{equation}
where $F_{V}(\Delta_{i})$ is the fraction of the volume with $\Delta <
\Delta_{i}$, and $\lambda_{o} H= 60 \, {\rm km s^{-1}}$ (basically
determined by the Jeans length) is a good fit to numerical simulations.  

There is only a finite stretch of the spectrum over which OI and SiII
absorption can be seen before it becomes confused with the hydrogen
Ly$\alpha$ forest. Suppose we observe a bright quasar which
ionizes its surroundings so that the damping wing of the Gunn-Peterson
trough does not extend redward of the Ly$\alpha$ line. A photon can
redshift for $l = \delta \lambda/\lambda \approx 20,000 \, {\rm km \,
s^{-1}}$ from the OI absorption edge at $1302(1+z_{s}) {\rm \AA}$ and $l \approx 10,000 \, {\rm km \,
s^{-1}}$ from the SiII absorption edge at $1260(1+z_{s}) {\rm \AA}$, before it encounters the HI
Gunn-Peterson trough (this interval is actually somewhat smaller, due
to the finite width of Ly$\alpha$ and NV emission lines). The average number
of lines with $N > N_{i}$ is $\sim \lambda_{i}/l$, while the
probability that no lines with $N > N_{i}$ are seen is
exp($-\lambda_{i}/l$). 

In Figure \ref{fig:pre_overlap}, we show the mean number of OI lines above
a given column density $\tilde{N}_{\rm OI}$ detectable in the pre-overlap
phase. In this phase, $(\bar{Z},f_{Z})$ are free parameters. For a
given $f_{Z}$, the column densities can be rescaled to the assumed
mean metallicity via $N_{\rm OI}=\tilde{N}_{\rm OI} \metallicity$. We
see that $\sim$few lines can be seen in the $10^{14}-10^{15} {\rm cm^{-2}}$ range in the pre-overlap era for
filling factors $f_{Z} \sim 10\%$. Somewhat more lines can be seen at higher
redshift since $N_{i} \propto (1+z)^{1/2}$. The two distinct slopes in
the relation can be easily understood. For a given $\bar{Z}$, increasing the metal-filling
factor $f_{z}$ increases the number of patches along a line of sight
to a quasar which are metal-polluted, but decreases their mean
metallicity $Z_{\rm crit}$. For a low column density threshold, the
former effect dominates and $N_{\rm lines}$ increases with $f_{Z}$;
for a high column density threshold, increasing $f_{Z}$ increases the
number of lines which fall below threshold, and $N_{\rm lines}$
decreases with $f_{Z}$. 

\begin{figure}
\psfig{file=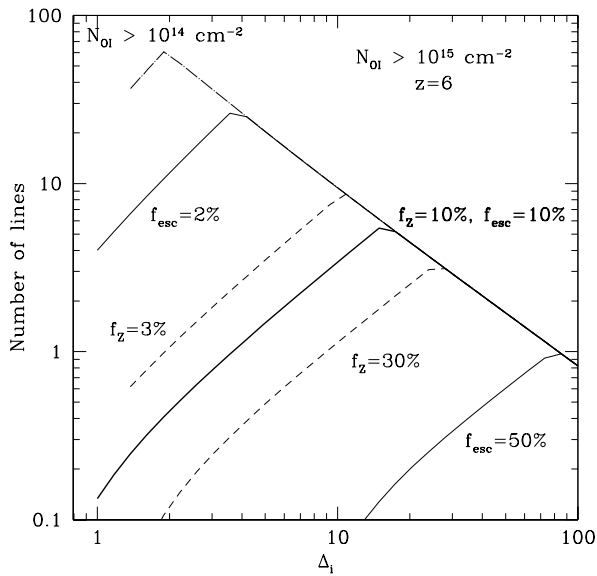,width=80mm}
\caption{Number of OI lines with $N_{OI} > 10^{15} {\rm cm^{-2}}$ observable in the post-overlap era at $z=6$ as a
function of $\Delta_{i}$, the overdensity to which the IGM is assumed to be
ionized. The fidicial model (shown in bold) is for
$f_{Z}=10\%,f_{\rm esc}=10\%$ (for a substantial QSO contribution,
$f_{\rm esc}$ should be replaced with $f_{\rm esc}/(1-f_{\rm
QSO})$). Other lines show the effect of varying $f_{Z},f_{\rm esc}$
and $N_{OI}$. The break in the slope of the relation can be easily
understood; see text for details. The number of observable SiII lines
with comparable equivalent widths is roughtly half that of OI.}
\label{fig:post_overlap}
\end{figure}

Similarly, in Fig. \ref{fig:post_overlap} we plot the number of
lines which can be seen in the post-overlap era at redshift $z=6$.
The relation between $\bar{Z}$ and $\Delta_{i}$
shown in Fig. \ref{fig:overdensity_zbar}
has been assumed. The different lines illustrate the effect of varying
the model parameters around the fidicial model $(f_{\rm esc},f_{\rm
Z},N_{\rm OI}^{\rm crit})=(0.1,0.1,10^{15}{\rm cm^{-2}})$. Again, the
change in slope can be easily understood: as the ionized overdensity
$\Delta_{i}$ increases the number of lines seen initially increases, because of the larger
implied star-formation rate and thus higher $\bar{Z}$. At some point
the decrease in filling factor of neutral regions overwhelms the increase in metallicity, and the
number of lines decreases. The overdensity $\Delta_{i}$ at which this
break occurs depends on $f_{\rm esc}/(1-f_{\rm QSO})$ and $f_{Z}$, since
these parameters control the relationship between $\Delta_{i}$ and
$Z_{\rm crit}=\bar{Z}/f_{Z}$. For instance, the universe is ionized up to a much higher
$\Delta_{i}$ for a given $\bar{Z}$ if $f_{\rm esc}/(1-f_{\rm QSO})$ is
large. We see
that it is quite plausible for us to see OI absorption lines with
observed equivalent widths $W_{\lambda} \approx 5.3 \, \AA \left(
\frac{N_{i}}{10^{15} \, {\rm cm^{-2}}} \right) \redshift \, {\rm \AA}$ in the
SDSS $z=6.28$ quasar, which lies just at the tail end of reionization
$\Delta_{i} \sim$few.   

Interesting (though model-dependent) constraints on $\bar{Z}, f_{\rm
esc}, f_{\rm QSO}$ and $f_{\rm Z}$ might be possible from measurements of the OI
forest. For instance, the length of the dark region
and upper limit on the transmitted flux in the Ly$\alpha,\beta$ troughs give
constraints on $\Delta_{i}$ for a given model of structure formation;
\scite{fan01b} find for their $z=6.28$ quasar that $\Delta_{i} \sim
3$. From Figure \ref{fig:overdensity_zbar} we see that this implies
$\bar{Z}\approx 10^{-2.4} Z_{\odot} \left( \frac{f_{\rm esc}}{0.1}
\right) \left( \frac{1-f_{\rm QSO}}{1} \right)^{-1}$, fairly high
metallicities. From Figure \ref{fig:post_overlap} the number of OI
lines above a given column density in the post-overlap phase depends on
$f_{Z},f_{\rm esc}$; the observed number might constrain their value. The universe is likely to be in the pre-overlap era if a very long Gunn-Peterson
trough with no detectable flux is seen. From Figure
\ref{fig:post_overlap}, the number of detectable lines above a given
column density might constrain $\bar{Z},f_{Z}$. The numerical value of
the constraints are of course very model-dependent and should not be
over-interpreted. Still, some interesting statements might still be made with reasonable confidence. For instance, if no OI lines can be
seen in the post-overlap era despite a deep Gunn-Peterson damping
trough, then $\bar{Z}$ is low: this implies the escape fraction of
ionizing photons is close to unity and/or
quasars are the dominant ionizing source. Alternatively, the majority
of the metals are highly ionized, either because most metals reside
in voids (although this is unlikely given the finite speed at which
metal pollution fronts can propagate), or the high-density regions in
which they reside are constantly illuminated by an ionizing source.  

The main observational obstacle to detecting the OI forest is
confusion with other sources of line absorption. Intrinsic absorption within the quasar host can be
constrained by the relative absence of absorption features in the rest frame
$1216-1302 {\rm \AA}$ range of the large sample of lower redshift
quasars. Indeed, because of the reasonably wide wavelength interval in
which OI and SiII absorption can be seen ($\sim 20,000 {\rm km \, s^{-1}}$
and $\sim 10,000 {\rm km \, s^{-1}}$ respectively), at the shorter
wavelengths within this range intrinsic absorption can be ruled out. Broad
absorption line (BAL) quasars could have winds which show such
features, though usually the outflows are highly ionized. Confusion
with long wavelength metal lines such as CIV or MgII from lower-redshift
systems is another problem. Such an origin can be constrained by the
absence of a damped-Ly$\alpha$ system at the corresponding redshift in
the low redshift Ly$\alpha$ forest. The third and potentially most
serious problem is the fact that the night sky becomes increasingly
noisy at these near-IR wavelengths, and the atmospheric OH forest
becomes important. The accuracy with which the OI forest can be
detected therefore depends on the accuracy with which telluric
features can be divided out via a standard-star
calibration. Also, there are stretches between the night sky
OH forest lines in which no absorption should be seen, so OI
absorption can be identified if it falls within these regions. Ultimately, the presence of neutral gas can be corroborated
with simultaneous detections of OI, SiII and possibly FeII
absorption features. A larger quasar sample which shows long stretches of
complete HI Gunn-Peterson absorption and in which the density of OI
and SiII absorption features is higher in higher redshift quasars
should be an unambiguous signature of almost fully neutral patches of gas at high redshift.        

\section{Discussion}

The SDSS 1030+0524 quasar at $z=6.28$ shows tantalizing absorption
features blueward of the OI 1305${\rm \AA}$ line \cite{becker01}. There
also appears to be a fairly deep absorption feature blueward of the
SiII 1260$\AA$. Could the absorption lines described in this paper been
seen already? Unfortunately, the quasars SDSS 1044-0125($z=5.80$),
0836+0054($z=5.82$),1306+0356($z=5.99$) also show some absorption features
in the same wavelength interval; in particular, SDSS 1306+0356 shows a
very strong absorption feature at $\sim 7130 {\rm \AA}$, with no
detected flux over $\sim 80 {\rm \AA}$ (this has been tentatively
identified as CIV absorption at $z=4.86$). These features cannot
correspond to OI absorption lines: unless they correspond to regions of
anomalously high metallicity, the associated hydrogen column densities
would be $N_{HI} > 10^{20} {\rm cm^{-2}}$, and all flux at the
hydrogen Ly$\alpha$ wavelength should be obliterated, while some flux is
still seen there. These lines are probably associated with metal lines
(e.g. MgII) from lower redshift absorbers, and illustrate a generic
difficulty in observing the features proposed in this
paper. On the other hand, OI lines are not ruled out in the $z=6.28$
quasar because of the complete damping of flux at Ly$\alpha$,
Ly$\beta$ wavelengths. As we have seen, these absorption features can
still arise in the post-overlap epoch when regions with $\Delta > {\rm
few}$ are largely neutral; indeed, up to a few absorption lines with
observed equivalent widths $W_{\lambda} \sim 5 \redshift \left(
\frac{N_{i}}{10^{15} \, {\rm cm^{-2}}} \right)$ \AA $\,$ might be
seen. Note that the spectra of \scite{becker01} have been smoothed
to 4\AA $\, {\rm pixel}^{-1}$. The absorption features blueward of OI
1305${\rm \AA}$ cannot be definitely identified as OI absorption. Two
of them lie at the same wavelengths as bright sky emission lines and are probably due to
imperfect sky subtraction. A third line with observed frame equivalent
with $\sim 25 \, \AA$ is a possible candidate; however, it lies very
close to the rest frame OI wavelength and could be due to intrinsic
absorption. More cannot be said without careful
study of the spectrum. A definitive detection of the OI forest can
probably only be done with much higher signal-to-noise spectra of the
same quasar.   
     
The estimates in this paper can be addressed with in much
greater detail with numerical
simulations. In particular, I used very simple ansatzes for the
dependences of metallicity and ionization fraction with overdensity,
$Z(\Delta), x_{i}(\Delta)$ which in fact should be highly stochastic
and spatially varying. They can be much better modelled in a
self-consistent fashion in simulations which attempt to model the
metal pollution \cite{cenostriker,anthony} and radiative transfer
\cite{gnedin,raz}, particularly since the rise in metallicity and
ionization fraction are inter-related. The spatial structure of the OI forest
can be computed by shooting lines of sight through a simulation
box. If the OI forest is indeed seen, such studies will be urgently
needed to provide a more realistic interpretation of the
observations. 

The scenario in this paper is not significantly altered if a
substantial X-ray background due to high redshift supernovae
\cite{oh2001} or quasars \cite{venkatesan} is
present. The large mean free path of hard photons means that they can
ionize the IGM fairly uniformly, but beyond $x_{e} \sim 0.1$ most of the energy
of an energetic electron created goes into Coulomb heating the
gas rather than collisional ionization \cite{shull}; predominantly neutral
regions should therefore still exist.  

We will gain a wealth of information about early
metal pollution and the reionization process if the OI
and SiII forests are seen. They will be direct probes of the topology
and history of gas clumping, metal pollution and reionization in the
early universe. As we have seen, if we assume a relation between the
metals and ionizing photons produced by massive stars, they could
potentially also provide indirect constraints
on the escape fraction of ionizing photons from star forming halos and
the QSO contribution to the ionizing background, and the filling
factor of metal pollution. They may also give clues as to the nature of the ionizing sources: the structure of forest lines
should look different if the universe were reionized by rare but
luminous source as opposed to abundant but faint sources, since in the
former case higher overdensity regions have to be ionized before
overlap can be achieved. OI and HI will be locked in very tight charge
exchange equilibrium at high redshift. If we are lucky enough to observe the
rest-frame optical afterglow of a high-redshift gamma-ray burst and
measure both $\tau_{\rm HI}^{\rm eff}$ (from the damping
wind) and $\tau_{\rm OI}^{\rm eff}$, we will have a direct
measure of the mean metallicity of the universe at high redshift,
independent of gas clumping or the form of the ionizing radiation
field. Otherwise, a lower limit on the metallicity from measurement of
$\tau_{\rm OI}^{\rm eff}$ alone is possible. A null detection of the
OI, SiII forests will yield constraints on the parameter $Y_{i}(\Delta)= x_{i}(\Delta)
Z(\Delta)$, but a positive detection will be tremendously exciting and
almost certainly signal the presence of almost fully neutral hydrogen
at high redshift. To date, the OI and SiII forests may be our
only probes of nearly neutral gas in the pre-reionization epoch
observable with current technology. 

\section*{Acknowledgements}

I am very grateful to Michael Strauss for detailed comments, and
Charles Steidel for an informative conversation on some of the
observational issues. I also thank Xiaohui Fan, Avi Loeb and 
David Spergel for stimulating conversations, and Marc Kamionkowski for
encouragement and advice. I acknowledge NSF grant AST-0096023 for support.

\end{document}